\documentclass[a4paper,11pt]{article}
\usepackage{pos}
\usepackage{amsmath}
\usepackage{cleveref}
\usepackage{physics}
\usepackage{microtype}
\usepackage{subcaption}
\usepackage{placeins}
\hypersetup{
  pdftitle={Global analyses of PDFs in the precision era},
  pdfauthor={Juan M. Cruz-Martinez}
}

\graphicspath{{figures/}}

\title{Global analyses of PDFs in the precision era}
\ShortTitle{Global analyses of PDFs}

\author*[a]{Juan M. Cruz-Martínez}

\affiliation[a]{Universidad de Sevilla, Departamento de Física Atómica, Molecular y Nuclear,\\
  Facultad de Física, Universidad de Sevilla, E-41080 Sevilla, Spain}

\emailAdd{jcruz@us.es}

\abstract{
    Parton distribution functions or PDFs are determined through global fits to an ever-increasing array of data, including deep-inelastic scattering at fixed-target experiments and HERA, and hadron collisions at the Tevatron and LHC. As the dataset has expanded, theoretical accuracy has improved from partial next-to-next-to-leading order (NNLO) coverage about 20 years ago to approximate N$^3$LO global fits today. The fitting collaborations must reconcile these complex calculations and data analyses while ensuring that quoted PDF uncertainty bands remain reliable.
}

\FullConference{The 33rd International Workshop on Deep Inelastic Scattering and Related Subjects (DIS2026)\\
4--8 May 2026\\
Bologna, Italy}

\begin{document}
\maketitle

\section{PDFs as a precision interface}

As Run III of the Large Hadron Collider (LHC) comes to an end in 2026, the increased precision made possible by the data collected since the start of its physics programme in 2010 has opened new opportunities to stress-test the Standard Model by searching for deviations from its predictions.
The experimental effort to reduce uncertainties through the sheer amount of data, careful analyses and detailed knowledge of the instrumental setup must be accompanied by a similar effort from the theoretical community, especially in the calculation of higher-order perturbative corrections.
In quantum chromodynamics (QCD), predictions for many important $2 \to 1$ and $2 \to 2$ processes are known and are publicly available at next-to-next-to-leading order (NNLO) (i.e., two powers of $\alpha_s$ beyond the leading-order or LO contribution).
For key processes, such as inclusive Drell--Yan production and Higgs-boson production, predictions are available at one order higher, N$^3$LO~\cite{Baglio:2022wzu}. See Refs.~\cite{Groeber:2025bsm,Alekhin:2024mrq,Huss:2025nlt} and references therein for more complete discussions of the status of precision calculations.

This is, however, only part of the story.
The precision of theoretical predictions also depends on our knowledge of the internal structure of the proton itself.
Its longitudinal partonic structure is encoded in the so-called parton distribution functions (PDFs), and their accurate determination is one of the essential parts of today's phenomenological efforts, even more so in preparation for the upcoming High-Luminosity upgrade of the LHC (HL-LHC).

\begin{figure}[h]
	\centering
	\includegraphics[width=0.85\textwidth]{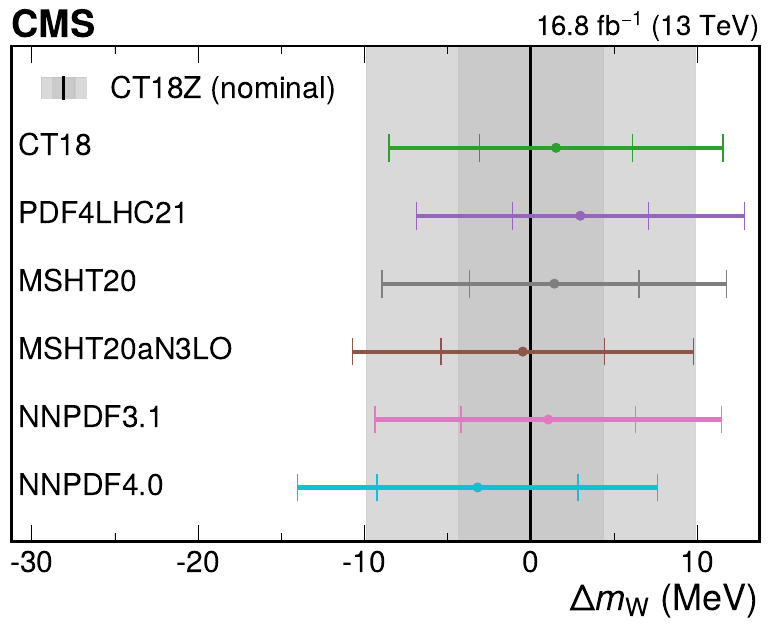}
    \caption{Difference in the $W$-boson mass extracted by CMS using different PDF sets, relative to the nominal result (CT18Z). For some sets, input PDF uncertainties are rescaled before refitting. The inner bars show the PDF uncertainty, while the outer bars show the total uncertainty. The PDFs are profiled by CMS, which modifies their nominal uncertainties. Figure taken from Ref.~\cite{CMS:2024lrd}. }
	\label{fig:wmass}
\end{figure}

A paradigmatic example of the above is the recent extraction of the $W$-boson mass by CMS~\cite{CMS:2024lrd}.
In this determination, the post-fit PDF uncertainty impact is 4.4~MeV on the mass of the $W$, compared with a total uncertainty of 9.9~MeV, making it one of the dominant sources of uncertainty.
This can be seen graphically in~\cref{fig:wmass}, which compares the mass values extracted with several PDF sets~\cite{Hou:2019efy,Bailey:2020ooq,NNPDF:2017mvq,NNPDF:2021njg,PDF4LHCWorkingGroup:2022cjn,McGowan:2022nag}, with CT18Z~\cite{Hou:2019efy} as the baseline.
With such a direct impact, a careful estimate of PDF uncertainties is crucial for the physics programme of the LHC and for ensuring that improvements in experimental precision translate into stronger tests of the underlying physics.

PDFs, however, are nonperturbative quantities and cannot be calculated using perturbative QCD alone.
Although lattice QCD provides complementary information on proton structure, precision phenomenology still relies on their extraction from experimental data through global fits~\cite{Cichy:2019ebf,DelDebbio:2020cbz,Constantinou:2020hdm}.

The connection between the measured data and the PDFs is provided by collinear factorization.
For a sufficiently inclusive hard proton-proton cross section ($\mathcal{O}$) to which collinear factorization applies, this separates the long-distance proton structure from the perturbatively calculable partonic scattering, as shown schematically in~\cref{eq:inverseproblem}, neglecting power-suppressed contributions.

\begin{equation}
	\mathcal{O} = \sum_{i,j} \int_0^1 \dd x_1 \int_0^1 \dd x_2\,
	f_i(x_1,\mu_f)\,f_j(x_2,\mu_f)\,
	\hat{\sigma}_{ij}(x_1,x_2;\mu_r,\mu_f)\,,
	\label{eq:inverseproblem}
\end{equation}
where $f_i$ denotes the PDF for parton $i$ carrying a momentum fraction $x$ at factorization scale $\mu_f$, while $\hat{\sigma}_{ij}$ is the partonic cross section for the chosen observable and $\mu_r$ the renormalization scale.
PDF determination is thus an inverse problem: the measured cross sections constrain the unknown functions through these convolutions, with different processes probing different flavours and momentum fractions.

A PDF determination comprises three main components: data, theory and methodology.

\begin{figure}[h]
	\centering
	\includegraphics[width=1.0\textwidth]{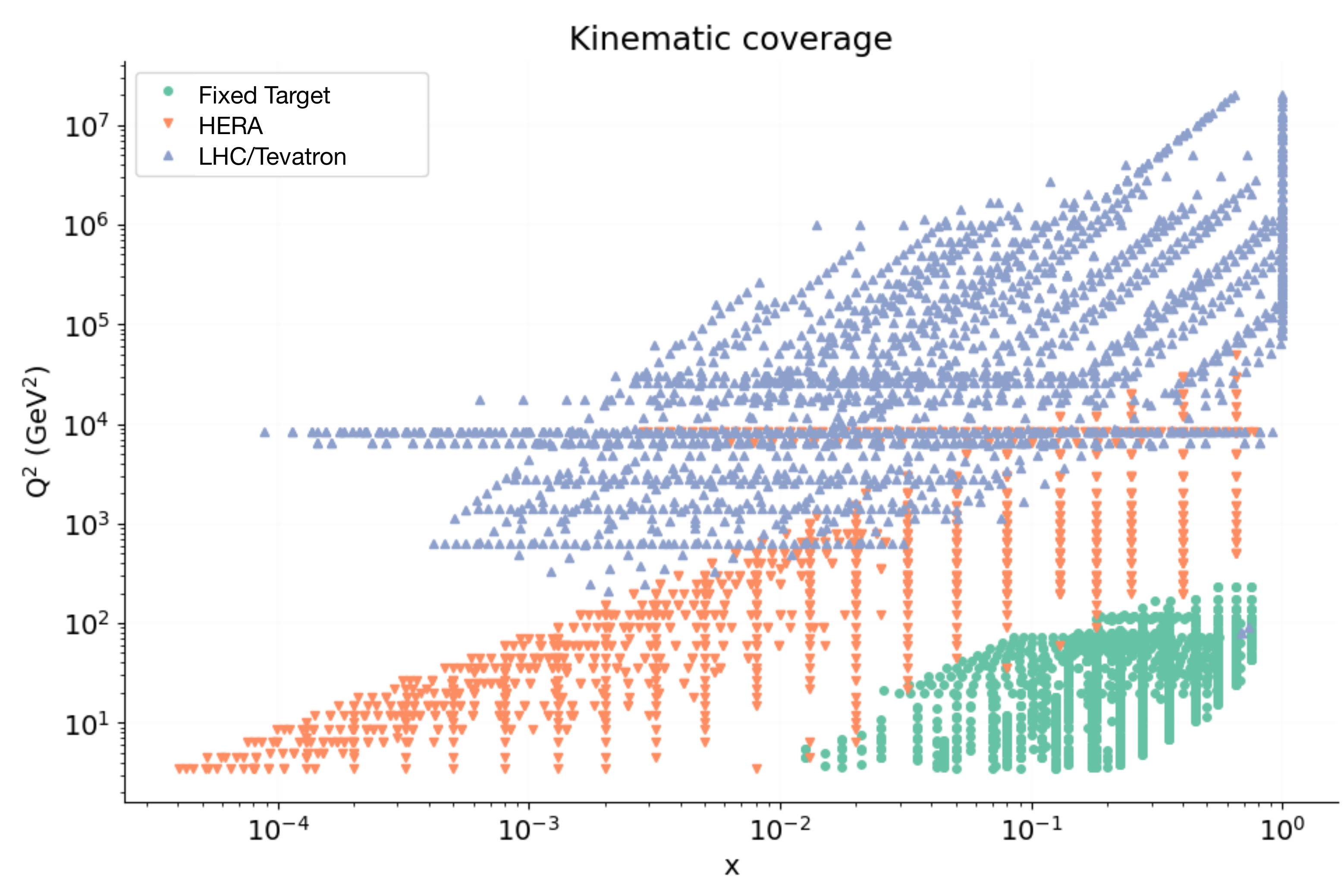}
	\caption{Kinematic coverage of the NNPDF4.0 data~\cite{NNPDF:2021njg} in the $x-Q^2$ plane. Fixed-target data include deep-inelastic scattering (DIS) and Drell--Yan production. HERA denotes collider DIS. LHC/Tevatron data include electroweak-boson, jet, $t\bar{t}$, single-top and prompt-photon production. For hadronic data, the plotted coordinates represent leading-order kinematics.}
	\label{fig:coverage}
\end{figure}

By data we mean the experimental measurements and subsequent analyses which provide us with the information needed to include them in a PDF determination, in particular their statistical and systematic uncertainties and correlations within and between measurements.
These are crucial for a faithful estimate of PDF uncertainties.
A detailed comparison of datasets included in recent determinations by NNPDF, MSHT, CTEQ and ABMP can be found in Appendix B of Ref.~\cite{NNPDF:2021njg}, with a global overview in~\cref{fig:coverage}. Beyond the datasets considered here, forward-neutrino measurements from FASER and future DIS data from the Electron–Ion Collider (EIC) offer new opportunities to constrain PDFs in complementary kinematic regions.

Theory includes, first and foremost, the perturbative accuracy of the predictions used in the fit and of the splitting function in the DGLAP evolution equations~\cite{Altarelli:1977zs,Gribov:1972ri,Dokshitzer:1977sg} that govern the scale dependence of the PDFs (today known at aN$^3$LO, see Refs.~\cite{Cridge:2025oel, Hekhorn:2025xke} for recent overviews).
We must not forget, however, that the theory entering a PDF determination also involves many choices: are electroweak effects accounted for? (and if so, how?) Is QED included in the PDF evolution alongside QCD~\cite{Cooper-Sarkar:2025sqw,Cridge:2026jyh}? Which exact values are chosen for the Standard Model parameters? Are resummation effects taken into account? Is the charm quark purely perturbative, or do we fit it independently, allowing for intrinsic charm~\cite{Guzzi:2022rca,Ball:2022qks}?
These and other choices define the theoretical framework under which the PDFs are determined and can potentially drive the difference between extractions as much as the data.

Finally, to put everything together, a methodology must be defined.
The methodology specifies how the PDFs are parametrized, how the parameters are optimized, and how uncertainties are propagated into the final result.
The parametrizations used in global fits can be roughly divided into two broad families: explicit functional forms with fitted coefficients, adopted by most groups, and neural-network representations, whose flexibility is intended to reduce parametrization bias.
In the case of MSHT and CTEQ, the former is paired with Hessian uncertainty propagation, based on the behaviour of the fit's $\chi^2$ around its minimum and an appropriate tolerance prescription~\cite{Hou:2019efy,Bailey:2020ooq}.
NNPDF, which uses neural networks, instead fits an ensemble of networks to Monte Carlo replicas of the data~\cite{NNPDF:2021njg}.

\section{The evolution of PDFs during the LHC runtime}

\begin{figure}[!ht]
	\begin{subfigure}[t]{0.49\textwidth}
		\centering
		\includegraphics[width=\textwidth]{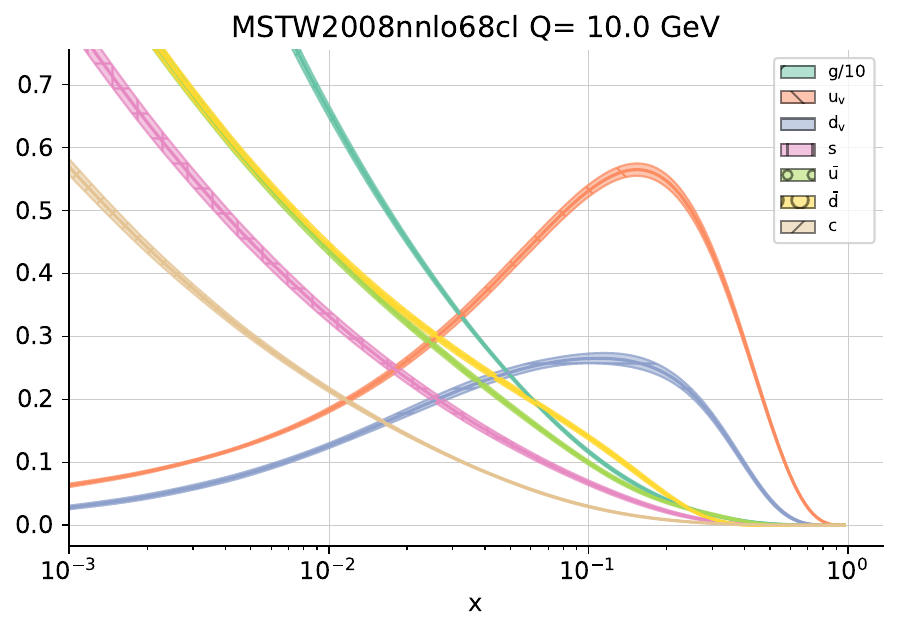}
		\includegraphics[width=\textwidth]{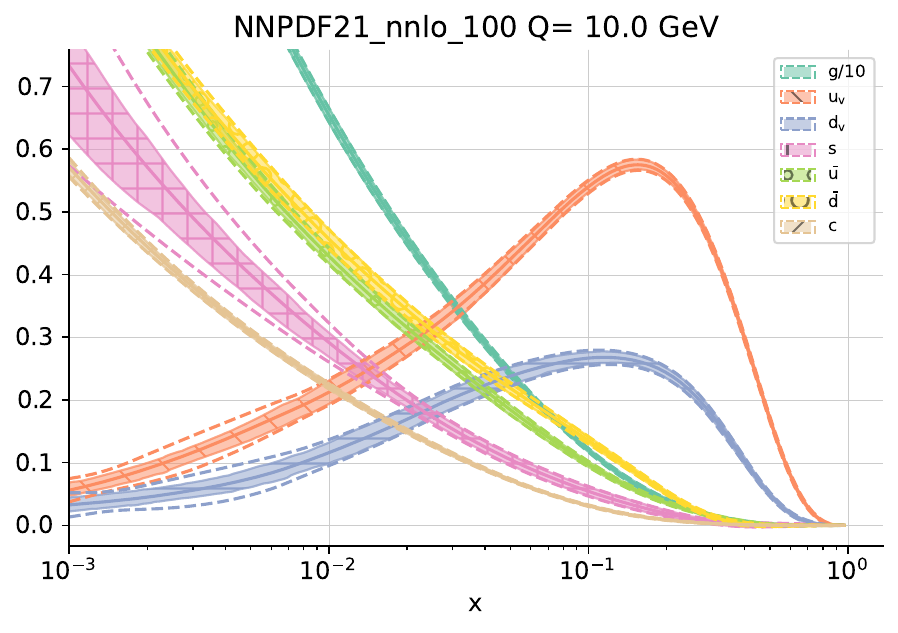}
		\includegraphics[width=\textwidth]{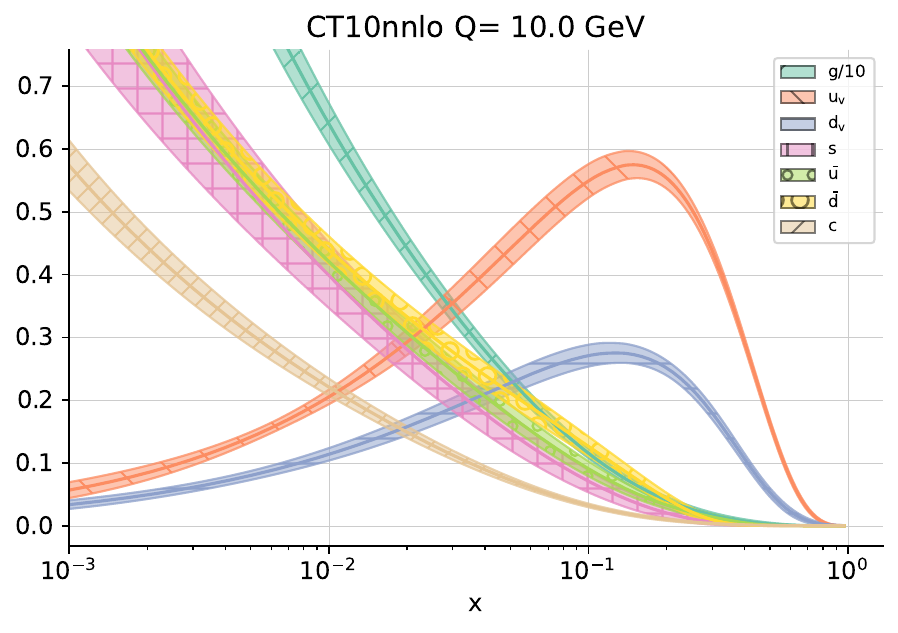}
		\caption{PDFs from around 2010.}
		\label{fig:pdfs_old}
	\end{subfigure} \hfill
	\begin{subfigure}[t]{0.49\textwidth}
		\centering
		\includegraphics[width=\textwidth]{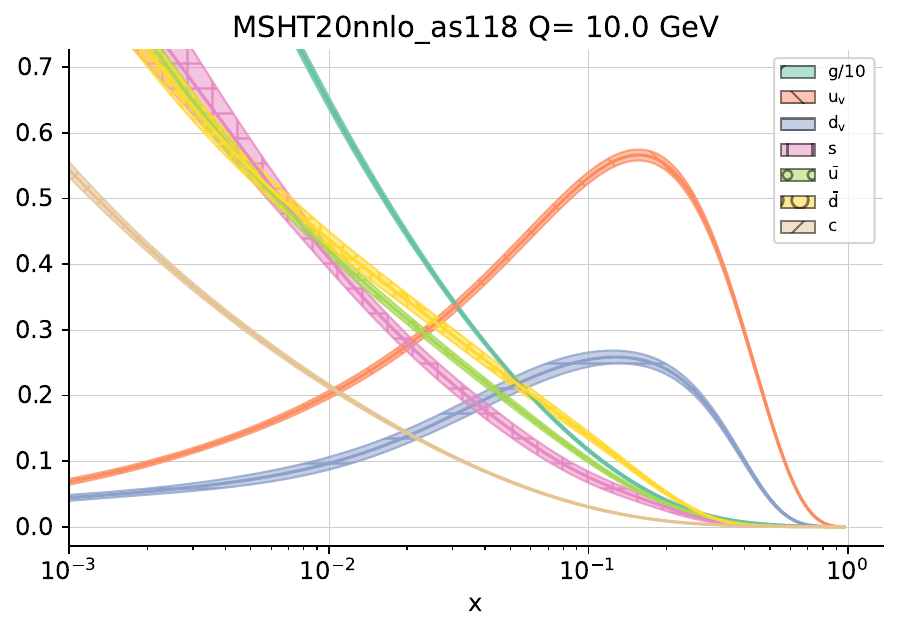}
		\includegraphics[width=\textwidth]{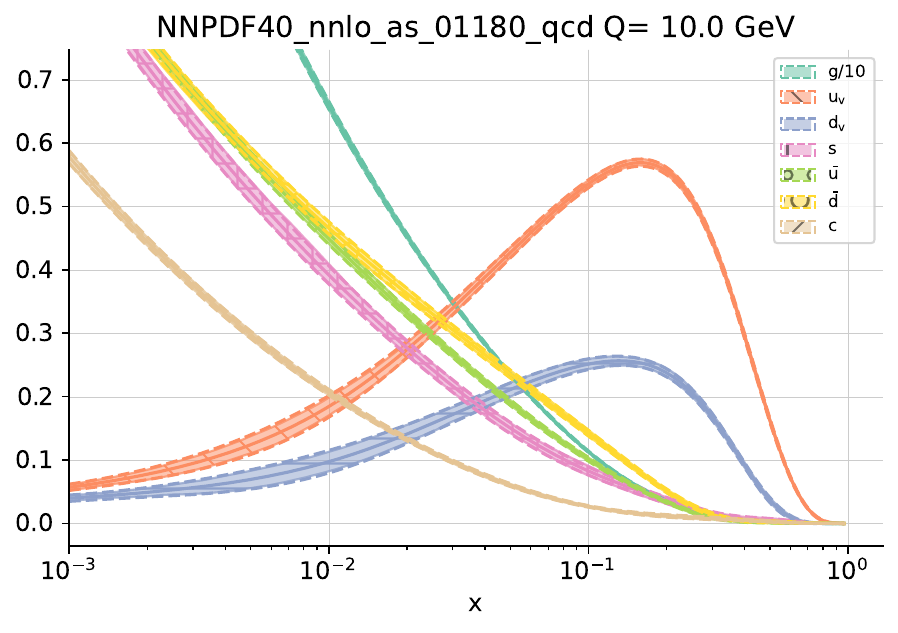}
		\includegraphics[width=\textwidth]{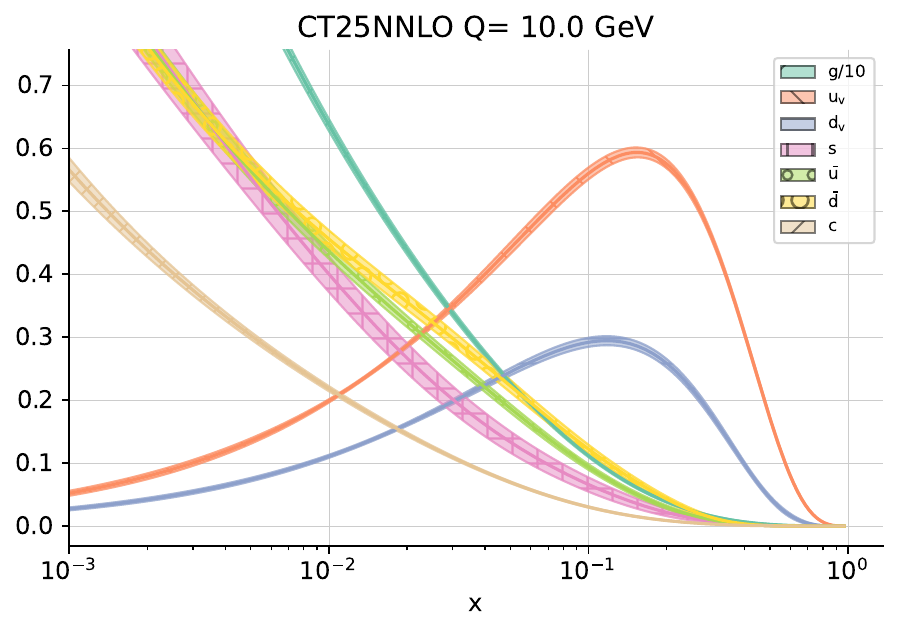}
		\caption{Recent PDFs, from 2020 onwards.}
		\label{fig:pdfs_new}
	\end{subfigure}
    \caption{Comparison of older PDF sets (showing $xf_{i}(x,Q)$) fitted without LHC data (left) and more recent sets including LHC measurements (right), from MSTW/MSHT, NNPDF and CTEQ, from top to bottom. The PDFs are evaluated at $Q=10$~GeV, with the gluon contribution divided by ten for visibility. The newer datasets extend the kinematic coverage and add constraints on several parton flavours.}
	\label{fig:all_pdfs_new_and_old}
\end{figure}

Over the sixteen years from the start of the LHC to the end of Run III, PDF determinations have undergone notable changes through new experimental measurements, more accurate predictions and updates to the fitting methodologies.
To keep this discussion brief, we focus on the determinations by the MSHT, CTEQ and NNPDF collaborations.
These are by no means the only PDF fitting collaborations, and global fits of unpolarized collinear PDFs represent only one part of the broader field of hadron structure.
These three groups contributed to the PDF4LHC21 combination~\cite{PDF4LHCWorkingGroup:2022cjn} and have remained active throughout the LHC era,
allowing us to follow the evolution of PDF determinations.
In~\cref{fig:pdfs_old}, we show sets fitted without LHC data: MSTW2008~\cite{Martin:2009iq} (which today has become MSHT), NNPDF2.1~\cite{Ball:2011uy}, and CT10~\cite{Gao:2013xoa}.
These fits used NNLO evolution and the available NNLO hard-scattering predictions and were thus nominally ``NNLO''.
In particular, only DIS and inclusive Drell--Yan predictions were available among those relevant for PDF determination (and could not be incorporated fully in all cases), so many processes still required lower-order predictions or approximate corrections.

In~\cref{fig:pdfs_new}, we show MSHT20~\cite{Bailey:2020ooq}, NNPDF4.0~\cite{NNPDF:2021njg} and the upcoming CT25~\cite{Ablat:2025gdb}.
These are still NNLO determinations, but they differ substantially from their predecessors.
First of all, many new measurements have been added, with LHC data providing important constraints in several kinematic regions (\cref{fig:coverage}).
The theoretical description at NNLO is also much more complete\footnote{For some processes, the computational cost of the NNLO predictions still motivates the use of pre-computed factors, defined as ratios of NNLO to NLO cross sections. In most cases this is just a technical hurdle and not a limitation on our current knowledge.}, making a more consistent treatment of the fitted data now possible.
Lastly, we must also consider the considerable methodological progress made during the LHC era.
With higher-order calculations reducing perturbative uncertainties in many observables and N$^3$LO predictions available for key channels such as inclusive Higgs and Drell--Yan production, PDF uncertainties increasingly limit precision phenomenology.
Any reduction in these uncertainties must therefore be understood and validated, given its substantial phenomenological implications.

\section{Beyond NNLO PDFs}

An important development in precision phenomenology is the extension of PDF determinations beyond NNLO, using N$^3$LO coefficients when available and increasingly complete information on DGLAP evolution.
The ingredients include massless DIS coefficient functions, selected hadron-collider cross sections, approximate N$^3$LO evolution and approximate heavy-quark contributions.
This information allows us to explore the PDFs beyond NNLO before every ingredient of a fully N$^3$LO global fit is available.

\begin{figure}[h]
	\centering
	\includegraphics[width=0.49\textwidth]{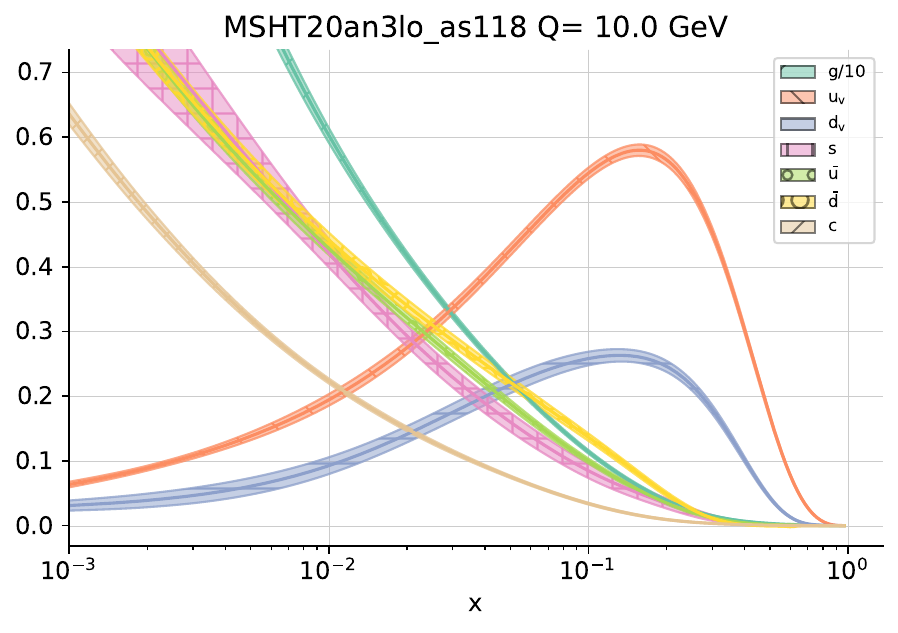}\hfill
	\includegraphics[width=0.49\textwidth]{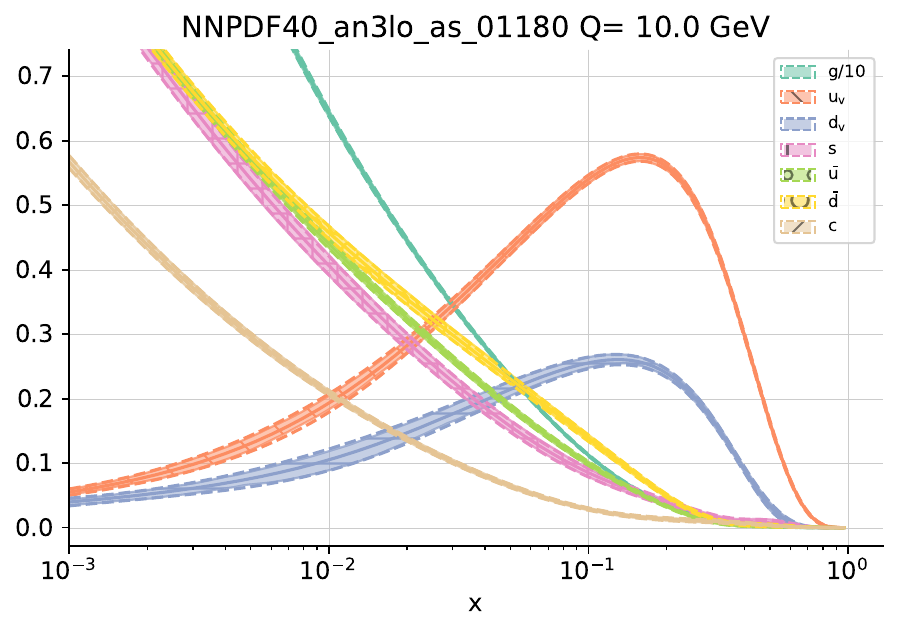}\hfill
    \caption{PDF determinations at (approximate) N$^3$LO by the MSHT (left) and NNPDF (right) collaborations, shown in the same format as in~\cref{fig:all_pdfs_new_and_old}.}
	\label{fig:flavs_n3lo}
\end{figure}

In~\cref{fig:flavs_n3lo}, we show the aN$^3$LO determinations by MSHT~\cite{McGowan:2022nag} and NNPDF~\cite{NNPDF:2024nan}.
The ``a'' in front stands for ``approximate'', indicating that these fits used the incomplete N$^3$LO information available.
Uncertainties on the incomplete ingredients were estimated using nuisance parameters in MSHT~\cite{Ball:2021icz,McGowan:2022nag} and covariance matrices in NNPDF~\cite{NNPDF:2019ubu,NNPDF:2019vjt,NNPDF:2024dpb}.
Although inclusive Drell--Yan predictions are already publicly available at N$^3$LO~\cite{Baglio:2022wzu} and fiducial calculations can be performed~\cite{Chen:2022cgv}, these do not yet provide all the corrections needed for the hadron-collider data entering these fits.
The cited determinations therefore use NNLO hadronic predictions supplemented by approximate corrections or additional uncertainties.
A comparison of the approximate N$^3$LO evolution implementations used by several PDF fitting groups is presented in Ref.~\cite{Cooper-Sarkar:2024crx}.
For phenomenological applications, Ref.~\cite{Cridge:2024icl} provides a combination of the MSHT and NNPDF aN$^3$LO sets.
The four-loop non-singlet splitting functions are now known in full~\cite{Gehrmann:2026qbl}.
Singlet approximations can be constructed from calculated Mellin moments and endpoint constraints~\cite{Falcioni:2025hfz}.
Note the fits in~\cref{fig:flavs_n3lo} predate these developments.
The respective papers detail the information included in each fit.

\begin{figure}[h]
	\centering
	\includegraphics[width=0.49\textwidth]{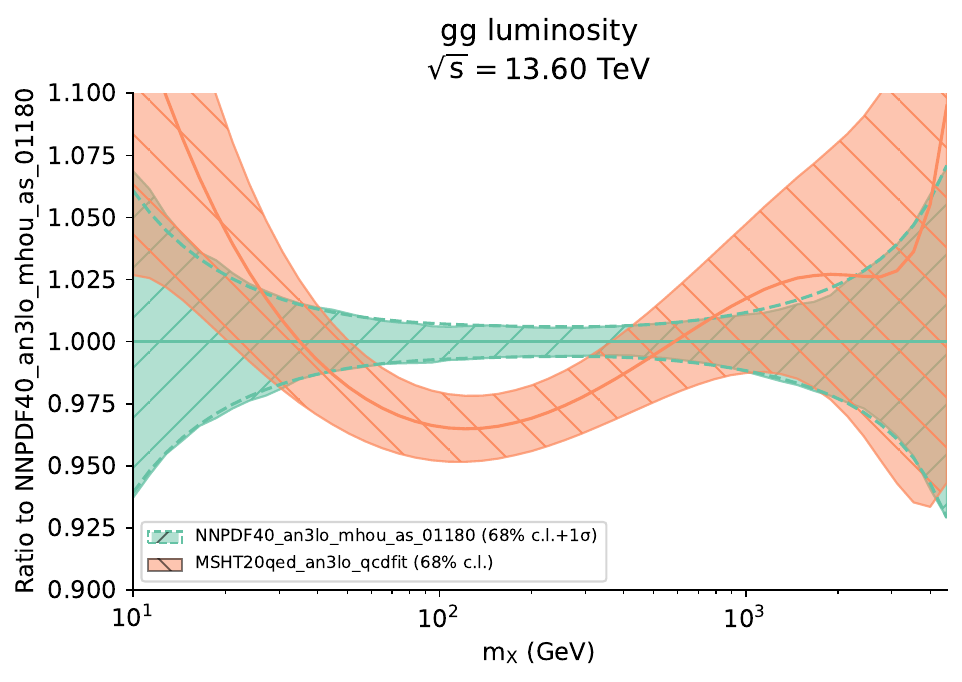}\hfill
	\includegraphics[width=0.49\textwidth]{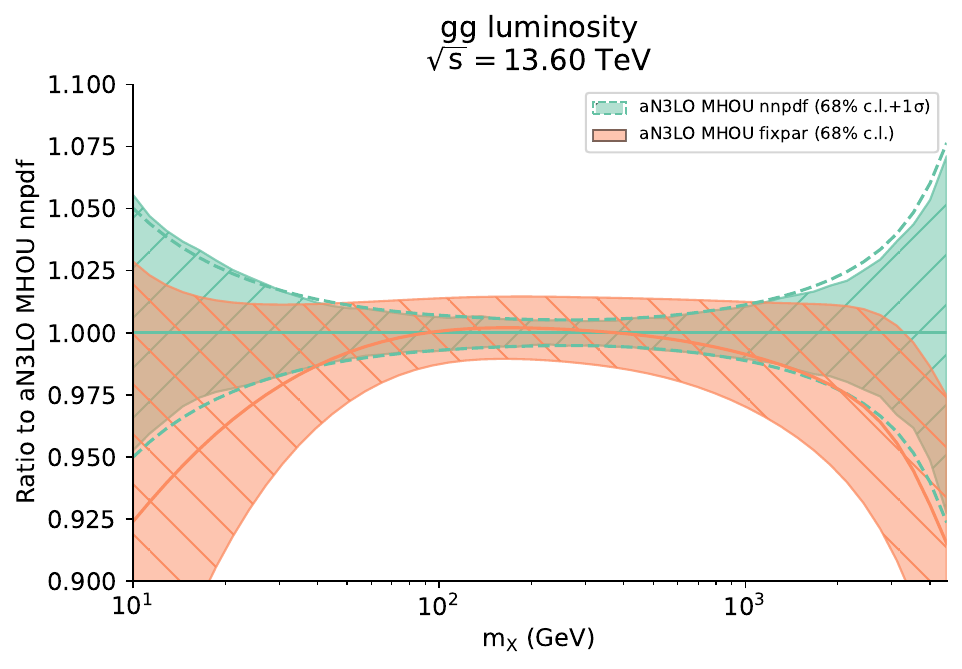}\hfill
        \caption{Gluon-gluon luminosities from published MSHT and NNPDF aN$^3$LO sets (left) and from fits using common data and theory inputs (FPPDF fit taken from Ref.~\cite{Cruz-Martinez:2026rct}). By fitting exactly the same data with the same theory, we can isolate the effects of the fitting methodology. In this case, for example, previously tensioned results become compatible, indicating the discrepancy was not driven by the methodology.}
	\label{fig:fppdf}
\end{figure}

At this level of accuracy, the community's attention has shifted towards precision: validating the PDF uncertainties is as important as reducing them.
Historically, these uncertainties have largely reflected the propagation of experimental errors into the PDFs.
In the Monte Carlo approach, one fits an ensemble of pseudodata replicas, typically sampled from a multivariate Gaussian distribution encoding the experimental uncertainties and correlations.
In the Hessian approach, one expands the fit's $\chi^2$ around its minimum, with a criterion of $\Delta\chi^2=1$ giving the 1$\sigma$ uncertainty intervals.
To account for departures from ideal statistical assumptions, including possible tensions between datasets, Hessian fits introduce the notion of tolerance~\cite{Collins:2001es,Pumplin:2001ct,Pumplin:2009sc,Martin:2009iq}, which enlarges the uncertainties by modifying the $\Delta\chi^2$ criterion through an appropriate prescription.
Note that, in the ideal scenario, when they describe the same Gaussian probability distribution, the Hessian and Monte Carlo methods are expected to agree~\cite{Watt:2012tq}.
Recently, FPPDF~\cite{Cruz-Martinez:2026rct}, built on the open-source NNPDF framework, has made it possible to benchmark fixed-parametrization/Hessian fits with their neural-network counterparts with common data and theory inputs, as illustrated in~\cref{fig:fppdf}.

A widely used tool for validating PDF uncertainties is the ``closure test'': fitting pseudodata generated from known PDFs with a specified theory and uncertainty model. Standard tests use mutually consistent pseudodata; controlled inconsistencies can also be introduced.
Such tests were explored in Ref.~\cite{Watt:2012tq} and became a central part of the validation of the NNPDF methodology~\cite{NNPDF:2014otw}.
They can be used to assess both the reconstruction bias and the statistical coverage of the uncertainties, i.e., whether quoted uncertainty intervals contain the true value with the expected frequency over repeated experiments.

Closure tests have since been studied in depth from statistical and practical perspectives, including tests with inconsistent pseudodata~\cite{DelDebbio:2021whr,Barontini:2025lnl}, and have recently also been adopted by the MSHT collaboration~\cite{Harland-Lang:2024kvt}.
A related body of work studies the statistical properties of PDFs in depth using alternative or explainable models~\cite{Kriesten:2023uoi,Candido:2024hjt,Kriesten:2024are,Costantini:2025wxp,Chiefa:2025cap,Barrue:2026qul,Bonnet-Guerrini:2026thq}.

\section{Conclusions}

During the LHC era, the three main ingredients of PDF determinations have progressed greatly:
large amounts of data, publicly available through HEPData, with carefully assessed and correlated uncertainties;
accurate fixed-order predictions, many of them publicly available through Monte Carlo generators, including N$^{3}$LO calculations;
and methodological advances that allow us to be both accurate and precise and to interpret the uncertainties we provide.
The overwhelming progress of this era gives us good reason to be optimistic about the future of PDF determinations for the HL-LHC.

\ \ 

\noindent This work is supported by MCIU/AEI/10.13039/501100011033 and FSE+ grant RYC2023-043794-I.

{\small
\setlength{\bibsep}{2pt}
\bibliographystyle{JHEP}
\bibliography{references}
}

\end{document}